\documentclass[runningheads]{llncs}
\usepackage[T1]{fontenc}
\usepackage{graphicx}
\usepackage{makecell, booktabs}
\usepackage[hidelinks]{hyperref}
\usepackage[capitalise]{cleveref}
\usepackage{color}

\begin{document}
\title{Robust and Generalisable Segmentation of Subtle Epilepsy-causing Lesions: a Graph Convolutional Approach }
\titlerunning{Segmentation of Subtle Epilepsy-causing Lesions}
%
\author{
Hannah Spitzer\inst{1,2}\orcidID{0000-0002-7858-0936} \and
Mathilde Ripart\inst{3}\orcidID{0000-0002-1761-5859} \and
Abdulah Fawaz\inst{4}\orcidID{0000-0002-7986-1801} \and
Logan Z. J. Williams\inst{4}\orcidID{0000-0001-5392-7043} \and
MELD project\inst{3} \and
Emma Robinson\inst{4}\orcidID{0000-0002-7886-3426} \and
Juan Eugenio Iglesias\inst{5,6,7}\orcidID{0000-0001-7569-173X} \and
Sophie Adler\inst{3}\orcidID{0000-0002-3978-7424} \and
Konrad Wagstyl\inst{8}\orcidID{0000-0003-3439-5808} 
}
\authorrunning{H. Spitzer et al.}
%

\institute{
Institute of Computational Biology, Helmholtz Center Munich, Munich 85764, Germany \and
Institute for Stroke and Dementia Research (ISD), University Hospital, LMU Munich, Munich, Germany \and
Department of Developmental Neuroscience, UCL Great Ormond Street Institute for Child Health, London WC1N 1EH, UK \and
Department of Biomedical Engineering, School of Biomedical Engineering and Imaging Sciences, King’s College London, London, UK \and
Martinos Center for Biomedical Imaging, MGH \& Harvard Medical School, USA \and
Computer Science and Artificial Intelligence Laboratory, Massachusetts Institute of Technology, Boston, 02139, USA \and
Centre for Medical Image Computing, University College London, London WC1V 6LJ, UK \and
Wellcome Centre for Human Neuroimaging, University College London, London WC1N 3AR, UK
}
\maketitle              
\begin{abstract}
Focal cortical dysplasia (FCD) is a leading cause of drug-resistant focal epilepsy, which can be cured by surgery. 
These lesions are extremely subtle and often missed even by expert neuroradiologists. 
“Ground truth” manual lesion masks are therefore expensive, limited and have large inter-rater variability. 
Existing FCD detection methods are limited by high numbers of false positive predictions, primarily due to vertex- or patch-based approaches that lack whole-brain context. 
Here, we propose to approach the problem as semantic segmentation using graph convolutional networks (GCN), which allows our model to learn spatial relationships between brain regions. 
To address the specific challenges of FCD identification, our proposed model includes an auxiliary loss to predict distance from the lesion to reduce false positives and a weak supervision classification loss to facilitate learning from uncertain lesion masks. 
On a multi-centre dataset of 1015 participants with surface-based features and manual lesion masks from structural MRI data, the proposed GCN achieved an AUC of 0.74, a significant improvement against a previously used vertex-wise multi-layer perceptron (MLP) classifier (AUC 0.64). 
With sensitivity thresholded at 67\%, the GCN had a specificity of 71\% in comparison to 49\% when using the MLP. 
This improvement in specificity is vital for clinical integration of lesion-detection tools into the radiological workflow, through increasing clinical confidence in the use of AI radiological adjuncts and reducing the number of areas requiring expert review.

\keywords{Graph Convolutional Network \and lesion segmentation \and structural MRI.}
\end{abstract}

\section{Introduction}
Structural cerebral abnormalities commonly cause drug-resistant focal epilepsy, which may be cured with surgery. Focal cortical dysplasias (FCDs) are the most common pathology in children and the third most common pathology in adults undergoing epilepsy surgery~\cite{Blumcke2017-xh}. 
However, 16-43\% of FCDs are not identified on routine visual inspection of MRI data by radiologists~\cite{Spitzer2022}. 
Identification of these lesions on MRI is integral for presurgical planning. 
Furthermore, accurate identification of lesions assists with complete resection of the structural abnormality, which is associated with improved post-surgical seizure freedom rates~\cite{Wagner2011-ud}.

There has been significant work seeking to automate the detection of FCDs, with the aim of identifying subtle structural abnormalities in patients with lesions not identified by visual inspection, termed “MRI negative”~\cite{Walger2023-ja}. 
These algorithms are increasingly being evaluated prospectively on patients who are “MRI negative” with suspected FCD, where radiologists review algorithm outputs and evaluate all putative lesions.
However, previous methods operate locally or semi-locally: using multilayer perceptrons (MLPs) which consider voxels or points on the cortical surface (vertices) individually~\cite{Spitzer2022,David2021-zb}, or convolutional neural networks which have to date typically been trained on patches of cortex~\cite{Gill2021-jf}. 
One widely-available algorithm using such an approach was able to detect 63\% of MRI negative examples, with an AUC of 0.64~\cite{Spitzer2022}. 
Overall, although these algorithms show significant promise in finding subtle and previously unidentified lesions, they are commonly associated with high false positive rates which hampers clinical utility~\cite{Walger2023-ja}. 
Detecting FCDs is particularly challenging due to small dataset sizes, high inter-annotator variability in manual lesion masks, and the large class imbalance, as FCDs typically only cover around 1\% of the total cortex. 
Nevertheless the urgent clinical need to identify more of these subtle lesions motivates the development of methods to address these challenges.

\subsubsection{Contributions}
\begin{figure}[t]
\includegraphics[width=\textwidth]{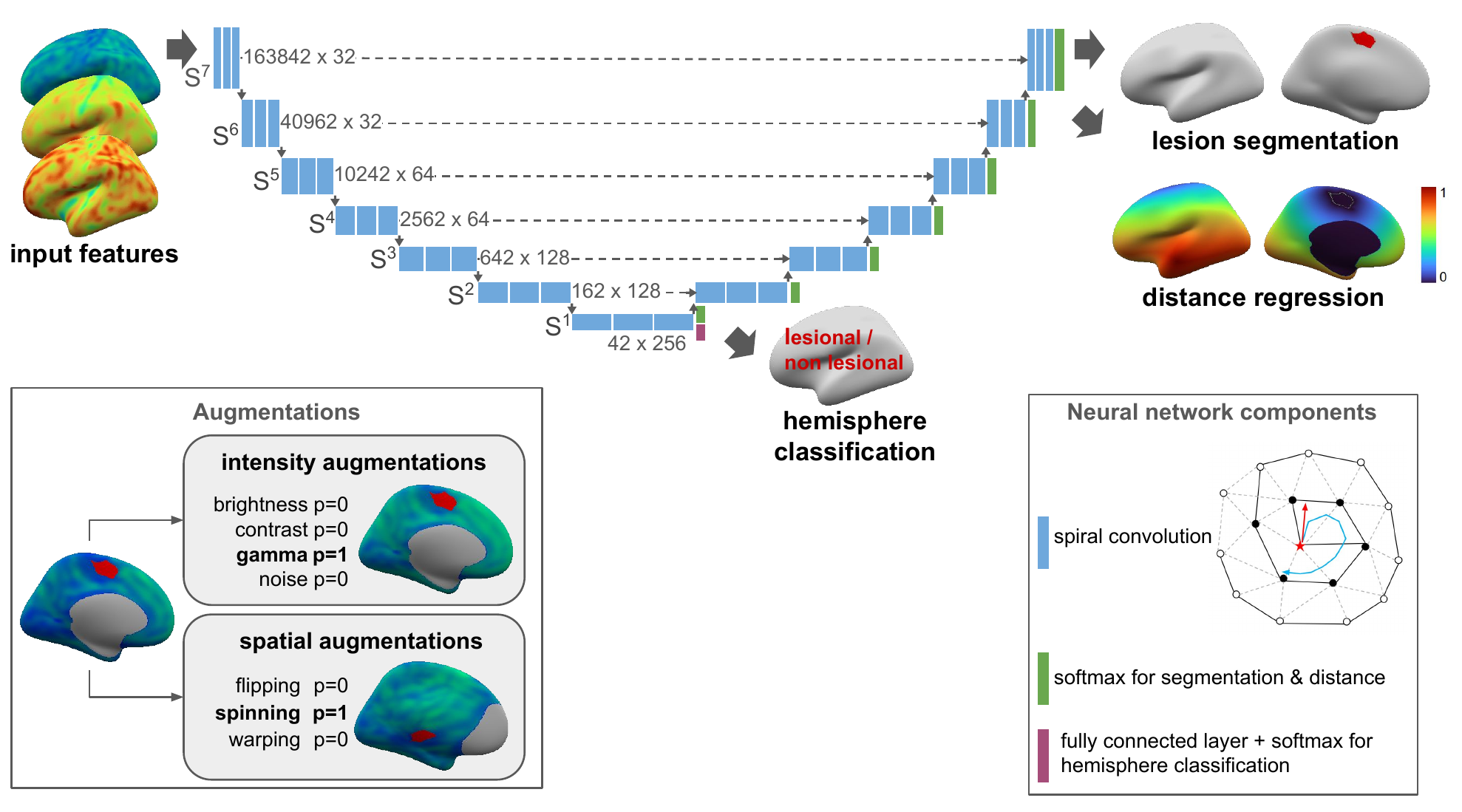}
\caption{
Proposed GC-nnU-Net+dc model for lesion segmentation, with auxiliary distance regression and hemisphere classification tasks. 
Lower left box: Types of data augmentation employed. 
Examples show the result of gamma intensity augmentation (top) and spinning (bottom).
} \label{fig1}
\end{figure}

We propose a robust surface-based semantic segmentation approach to address the particular challenges of identifying FCDs (\cref{fig1}). 
Our three main contributions to address these challenges are: 
1) Adapting nnU-Net~\cite{Isensee2021-yx}, a state-of-the-art U-Net architecture, to a Graph Convolutional Network (GCN) for segmenting cortical surfaces. 
This creates a novel method for cortical segmentation in general and for FCD segmentation in particular, in which the model is able to learn spatial relationships between brain regions. 
2) Inclusion of a distance loss to help reduce false positives, and 
3) Inclusion of a hemisphere classification loss to act as form of weak supervision, mitigating the impact of imperfect lesion masks. We directly evaluate the added value of each contribution on performance in comparison to a previously published MLP~\cite{Spitzer2022}. 
We hypothesised that the proposed GCN to segment FCDs would improve overall performance (AUC), in particular reducing the number of false positives (improved specificity) while retaining sensitivity. 
This improvement in classifier performance would facilitate clinical translation of automated FCD detection into clinical practice. 
All code to reproduce these results can be found at \href{github.com/MELDProject/meld_graph}{\texttt{github.com/MELDProject/meld\_graph}}.

\section{Methods}

\subsection{Graph convolutional network (GCN) for surface-based lesion segmentation}

We consider the lesion detection problem as a surface-based segmentation task. 
For this purpose, cortical surface-based features (intensity, curvature, etc.; see \cref{sub-dataset}) are extracted from each brain hemisphere and registered using FreeSurfer~\cite{Fischl2012-cx} to a symmetrical template. 
This template was generated by successively upsampling an icosahedral icosphere, $S^1$, with 42 vertices and 80 triangular faces. 
Icospheres $S^i$, with $i$ the resolution level of the icosphere, are triangulated spherical meshes, where $S^{i+1}$ is generated from $S^i$ by adding vertices at every edge. 
As input to our model we use icosphere $S^7$ (163842 vertices).

\subsubsection{U-Net architecture}
To segment lesions on the icosphere, we created a graph-based re-implementation of nnU-Net~\cite{Ronneberger2015-hd,Isensee2021-yx}. 
Unlike typical imaging data represented on rectangular grids, surface-based data require customised convolutions, downsampling and upsampling steps. 
Here, we used spiral convolutions~\cite{Gong2019-wg} which translates standard 2D convolutions to irregular meshes by defining the filter by an outward spiral. This ends up capturing a ring of information around the current node, similar to how a 2d filter captures a ring of information around the current pixel. We use a spiral length of 7, representing the central and adjacent 6 neighbours on a hexagonal mesh, roughly equivalent to a $3\times3$ 2D kernel. 
For downsampling from $S^{i+1}$ to $S^i$ in the U-Net encoder, a similar translation of 2D  max pooling is carried out by aggregating over all neighbours of the vertex at the higher-resolution $S^{i+1}$. 
Upsampling from $S^i$ to $S^{i+1}$ in the decoder is implemented via assigning the mean of each vertex in $S^i$ to all neighbours at level $S^{i+1}$. 
In total, the U-Net contains seven levels (mirroring the seven icospheres $S^7$-$S^1$), and every level consists of three convolutional layers using spiral convolutions and leaky Relu as activation function (\cref{fig1}). 

\subsubsection{Loss Functions}
Following best practices for U-Net segmentation models~\cite{Isensee2021-yx}, we use both cross-entropy and dice as loss functions for the segmentation, where $y$ is true labels, $\hat{y}$ is predicted, $n$ is the number of vertices:
\begin{equation}
L_{ce} = - \sum_{i=1}^{n} y_i \log(\hat{y}_i) + (1-y_i) \log(1-\hat{y}_i)
\end{equation}
\begin{equation}
L_{dice} = 1 - \frac{2 \sum_{i=1}^{n} y_i \hat{y}_i}{\sum_{i=1}^{n} y_i^2 + \sum_{i=1}^{n} \hat{y}_i^2 + \epsilon}
\end{equation}

\subsubsection{Distance loss}
To encourage the network to learn whole-brain context thereby reducing the number of false positives, we added an additional distance regression task. 
We train the model to predict the normalised geodesic distance $d$ to the lesion boundary for every vertex, by applying an additional loss $L_{dist}$ to the non-lesional prediction, $\hat{y}_{i,0}$ of the segmentation output for vertex $i$. 
We use the mean absolute error loss, weighted by the distance so as not to overly penalise small errors in predicting large distances from the lesion:
\begin{equation}
L_{dist}=\frac{1}{n} \sum_{i=1}^{n} \frac{| d_i - \hat{y}_{i,0} |}{d_i + 1}
\end{equation}

\subsubsection{Classification loss}
To mitigate uncertainty in the correspondence between lesion masks and lesions, we used a weakly-supervised classification loss $L_{class}$. 
For the ground truth $c$, examples were labelled as positive, if any of their vertices were annotated as positive. To predict this sample-level classification, we added a classification head to the deepest level (level 1) of the U-Net. 
The classification head contained a fully connected layer aggregating over all filters, followed by a fully-connected layer aggregating over all vertices, resulting in the classification output $\hat{c}$. 
This output was trained using cross-entropy:
\begin{equation}
    L_{class} = - \sum_{i=1}^{n} c_i \log(\hat{c}_i) + (1-c_i) \log(1-\hat{c}_i)
\end{equation}

\subsubsection{Deep supervision}
To encourage the flow of gradients through the entire U-Net, we use deep supervision at levels $I_{ds} = [6,5,4,3,2,1]$. Let $L_{ce}^i$, $L_{dice}^{i}$, $L_{dist}^{i}$ be the cross-entropy, dice and distance losses applied to outputs at level $i$, respectively. 
The model is trained on a weighted sum of all the losses, with $w_{ds}^i$ the loss weight at level $i$:
\begin{equation}
   L = L_{ce} + L_{dice} + L_{dist} + L_{class} + \sum_{i \in I_{ds}} w_{ds}^i ( L_{ce}^i + L_{dice}^i + L_{dist}^i) 
\end{equation}

\subsection{Data augmentation}
Data augmentations consisted of spatial augmentations and intensity augmentations (\cref{fig1}), following recommendations outlined in nnU-Net. Spatial augmentation included rotation, inversion and non-linear deformations of the surface-based data~\cite{Fawaz2021-ag}. 
Intensity-based augmentations included adding a Gaussian noise to the features intensity, adjusting the contrast, scaling the brightness by a uniform factor, and adding a gamma intensity transform. 

\section{Experiments and Results}

\subsection{Dataset and Implementation details}
\label{sub-dataset}

\subsubsection{Dataset}
For the following experiments, we used a dataset of post-processed surface-based features and manual lesion masks from 618 patients with FCD and 397 controls~\cite{Spitzer2022}. 
This is a heterogeneous, clinically-acquired dataset, collated from 22 international epilepsy surgery centres, including paediatric and adult participants scanned on either 1.5T or 3T MRI scanners. 
Each centre received local ethical approval from their institutional review board (IRB) or ethics committee (EC) to retrieve and anonymise retrospective, routinely available clinical data.
For each participant, MR images were processed using FreeSurfer~\cite{Fischl2012-cx} and 11 surface-based features (cortical thickness, grey-white matter intensity contrast, intrinsic curvature, sulcal depth, curvature and FLAIR intensity sampled as 6 intra- and sub-cortical depths) were extracted. 
FCDs were manually drawn by neuroradiologists to create 3D regions of interest (ROI) on T1 or fluid-attenuated inversion recovery (FLAIR) images. 
The ROIs were projected onto individual FreeSurfer surfaces and then the features and ROIs were registered to a bilaterally symmetrical template, fsaverage\_sym, using folding-based registration. 
Post-processing included 10mm full width at half-maximum surface-based smoothing of the per-vertex features, harmonisation of the data using Combat~\cite{Fortin2018-of} (to account for scanners differences), inter- and intra-individual z-scoring to account for inter-regional differences and demographic differences, and computation of the asymmetry index of each feature. 
The final surface-based feature set consisted of the original, z-scored and asymmetry features, resulting in 33 input features. 

In order to compare performance, the train/validation and test datasets were kept identical to those in the previously published vertex-wise classifier~\cite{Spitzer2022}. 
The train/validation cohort comprised 50\% of the dataset and 5-fold cross validation was used to evaluate the models. 
The remaining 50\% was withheld for final evaluation and comparison of models. 
Data from two independent sites (35 patients and 18 controls) were used to test the generalisability of the full model.

\subsubsection{Implementation details}
The graph-based convolutional implementation of nnU-Net (GC-nnU-Net) had the following training parameters: batch size: 8, initial learning rate: 10\textsuperscript{-4}, learning rate decay: 0.9, momentum: 0.99, maximum epochs: 1000 (1 epoch is 1 complete view of training data), maximum patience: 400 epochs; and augmentation probabilities: inversion: 0.5, rotation \& deformation: 0.2, Gaussian noise: 0.15, contrast: 0.15, brightness: 0.15, gamma: 0.15. Deep supervision weights were $w_{ds}=[0.5,0.25,0.125,0.0625,0.03125,0.0150765]$ for levels $I_{ds}=[6,5,4,3,2,1]$. 
Due to class imbalances, non-lesional hemispheres were undersampled during training to ensure 33\% of training examples contained a lesion. 
The model from the epoch with the best validation loss is stored for evaluation (Fig. S1).

Hardware: High-performance cluster with Single NVIDIA A100 GPU, 1000 GiB RAM; Software: PyTorch 1.10.0+cu11.1, PyTorch Geometric 2.0.4, Python 3.9.13. Combined memory footprint of model and dataset while training is 49GB.

\subsubsection{Experiments}
Using our graph-based adaptation of nnU-Net (GC-nnU-Net) and the previous MLP model as baseline, we ran an ablation study to measure the impact of the proposed auxiliary losses (\cref{tab1}). 
Each model was trained using the train/val cohort 5 times, withholding 20\% of the cohort for validation and stopping criteria. 
Final test performance was computed by ensembling predictions across the 5-fold trained models, with uncertainty estimates calculated through bootstrapping.
An additional experiment was carried out subsampling the training cohort at fixed fractions of 0.1, 0.2, 0.3, 0.4, 0.6 and 0.8 using the GC-nnU-Net+dc model (Fig. S2).

\begin{table}[t]
\centering
\caption{Experiments}\label{tab1}
\begin{tabular}{ll}
\toprule
\textbf{Experiment Name} & \textbf{Description} \\
\midrule
MLP~\cite{Spitzer2022} & vertex-wise multilayer perceptron \\
GC-nnU-Net & graph-based adaptation of nnU-Net \\
GC-nnU-Net+c & adding classification loss \\
GC-nnU-Net+d & adding distance loss \\
GC-nnU-Net+dc & adding distance loss and classification loss \\
\bottomrule
\end{tabular}
\end{table}

\subsubsection{Evaluations}
Model performances were compared according to their Area Under the Curve (AUC), which was calculated by computing the sensitivity and specificity at a range of prediction thresholds. 
For sensitivity calculations, due to uncertainty in the lesion masks, a lesion was considered detected if the prediction was within 20 mm of the original mask, as this corresponds with the inter-observer variability measured across annotators~\cite{Spitzer2022}. 
Specificity was defined by the presence of false positives in non-lesional examples. 
As an additional measure of model specificity, the number of false positive clusters in both patients and controls were calculated. 
Model AUCs were statistically compared using t-tests, with correction for multiple comparisons using the Holm-Sidak method.

\subsection{Results}
\begin{figure}[t]
\centering
\includegraphics[width=0.8\textwidth]{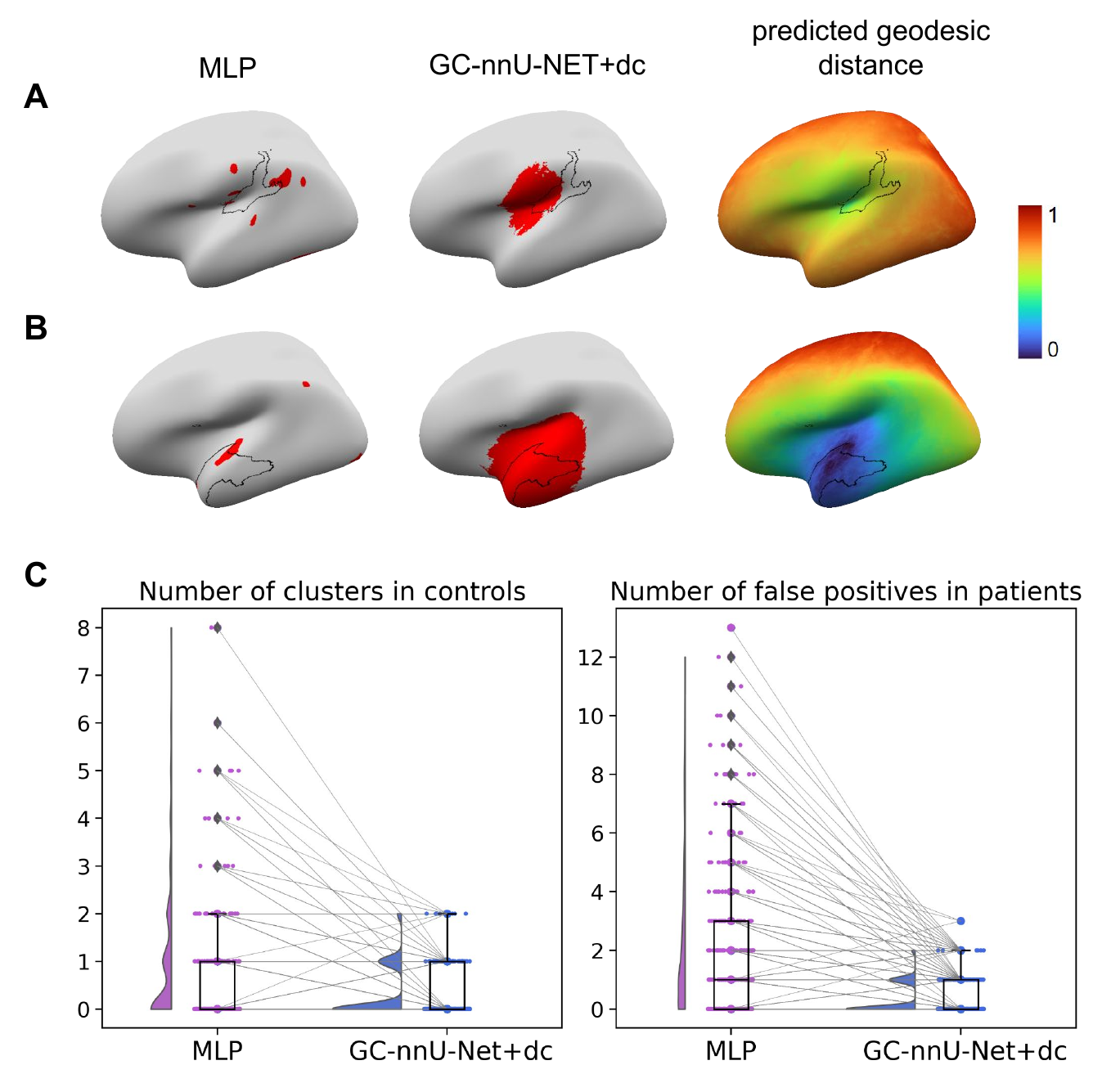}
\caption{
A \& B) Individual predictions for two patients using MLP and GC-nnU-Net+dc, as well as predicted geodesic distance from the lesion. 
Red: prediction. Black line: manual lesion mask.
Patients A and B both have false positive predictions using the MLP, unlike the predictions from GC-nnU-Net+dc.  
C) Comparison of number of clusters in controls and patients between MLP and GC-nnU-Net+dc on the test dataset. 
Grey lines: change in number of clusters for individual participants. 
} \label{fig2}
\end{figure}

\cref{tab2} compares model performances on the withheld test set. 
GC-nnU-Net+dc, the graph-based implementation of nnU-Net with additional distance and classification losses, outperformed all other models. 
Examples of individual predictions using the MLP and GC-nnU-Net+dc model, as well as examples of the predicted geodesic distance from the lesion are presented in~\cref{fig2}A,B. 
\cref{fig2}C visualises the reduction in number of false positive clusters when using GC-nnU-Net+dc which is reflected in the significantly improved specificity.
GC-nnU-Net+dc showed similarly improved specificity relative to the MLP on independent test sites, demonstrating good model generalisability (\cref{tab3}). 
In experiments varying the size of the training cohort, performance increased with sample size until around 220 subjects above which gains were negligible (Fig. S2).

\begin{table}[t]
\centering
\caption{Comparison of models on the test dataset.} \label{tab2}
\begin{tabular}{lllll}
\toprule
\textbf{Experiment} & \textbf{AUC (+/- std)} & \textbf{Sensitivity} & \textbf{Specificity} & \textbf{Run time (min)} \\
\midrule
MLP~\cite{Spitzer2022}    & 0.64 (n.a.)   & 67\%  & 49\%  & n.a.  \\
GC-nnU-Net                      & $0.68^{\dag\ddag}$ (+/- 0.004)    & 67\%  & 64\%  & 396.1 \\
GC-nnU-Net+c                    & $0.74^{\dag}$ (+/- 0.008)    & 67\%  & 66\%  & 373.9 \\
GC-nnU-Net+d                    & $0.69^{\dag\ddag}$ (+/- 0.007)    & 67\%  & 65\%  & 426.9 \\
GC-nnU-Net+dc                   & $0.74^\dag$ (+/- 0.005)           & 67\%  & 71\%  & 564.9 \\
\bottomrule
\end{tabular}
\vskip 3mm
\dag Model performance significantly improved compared to MLP. \\
\ddag Model performance significantly worse compared to GC-nnU-Net+dc.
\end{table}

\begin{table}
\centering
\caption{Comparison of models on independent test sites. } \label{tab3}
\begin{tabular}{lcccc}
\toprule
\textbf{Experiment} & \textbf{Sensitivity} & \textbf{Specificity} & \thead{\textbf{Median FP}\\ \textbf{in patients [IQR]}} &
\thead{\textbf{Median clusters} \\ \textbf{in patients [IQR]}}
  \\
\midrule
MLP~\cite{Spitzer2022}      & 79\%  & 17\%  & 2.0 [1.0, 4.0]  &  1.0 [1.0, 2.75]  \\
GC-nnU-Net+dc                     & 79\%  & 44\%  & 1.0 [0.0, 1.0]  &  1.0 [0.0, 1.0]  \\
\bottomrule
\end{tabular}
\vskip 3mm
FP: false positives, IQR: interquartile range.
\end{table}

\section{Conclusions and Future Work}
This paper presents a robust and generalisable graph convolutional approach for segmenting focal cortical dysplasias using surface-based cortical data. This approach outperforms specificity baselines by 22-27\%, which is driven by three newly-proposed components. 
First, treating the hemispheric surface as a single connected graph allows the network to model spatial context. 
Second, a classification loss mitigates the impact of imprecise lesion masks by simplifying the task to predicting whether or not a lesion is present in every hemisphere.
Third, a distance-from-lesion prediction task penalises false positives and encourages the network to consider the entire hemisphere. 
The results show a significant increase in specificity, both in terms of presence of any false positive predictions in non-lesional hemispheres and a reduced number of additional clusters in lesional hemispheres. 
From a translational perspective, this improvement in performance will increase clinical confidence in applying these tools to cases of suspected FCD, while additionally minimising the number of putative lesions an expert neuroradiologist would need to review. 
Future work will include systematic prospective evaluation of the tool in suspected FCDs and expansion of these approaches to multiple causes of focal epilepsy.

\subsubsection{Acknowledgements} 
The MELD project, MR and SA are supported by the Rosetrees Trust (A2665) and Epilepsy Research UK (P2208). KSW is supported by the Wellcome Trust (215901/Z/19/Z). LZJW is supported by the Commonwealth Scholarship Commission (United Kingdom). JEI is supported by the NIH (1RF1MH123195, 1R01AG070988, 1R01EB031114, 1UM1MH130981) and the Jack Satter Foundation. 

%
%
%
\bibliographystyle{splncs04}
\bibliography{references, references-anonymous}

\begin{thebibliography}{10}
\providecommand{\url}[1]{\texttt{#1}}
\providecommand{\urlprefix}{URL }
\providecommand{\doi}[1]{https://doi.org/#1}

\bibitem{Blumcke2017-xh}
Blumcke, I., Spreafico, R., Haaker, G., Coras, R., Kobow, K., Bien, C.G.,
  Pf{\"a}fflin, M., Elger, C., Widman, G., Schramm, J., Becker, A., Braun,
  K.P., Leijten, F., Baayen, J.C., Aronica, E., Chassoux, F., Hamer, H.,
  Stefan, H., R{\"o}ssler, K., Thom, M., Walker, M.C., Sisodiya, S.M., Duncan,
  J.S., McEvoy, A.W., Pieper, T., Holthausen, H., Kudernatsch, M., Meencke,
  H.J., Kahane, P., Schulze-Bonhage, A., Zentner, J., Heiland, D.H., Urbach,
  H., Steinhoff, B.J., Bast, T., Tassi, L., Lo~Russo, G., {\"O}zkara, C., Oz,
  B., Krsek, P., Vogelgesang, S., Runge, U., Lerche, H., Weber, Y., Honavar,
  M., Pimentel, J., Arzimanoglou, A., Ulate-Campos, A., Noachtar, S., Hartl,
  E., Schijns, O., Guerrini, R., Barba, C., Jacques, T.S., Cross, J.H., Feucht,
  M., M{\"u}hlebner, A., Grunwald, T., Trinka, E., Winkler, P.A., Gil-Nagel,
  A., Toledano~Delgado, R., Mayer, T., Lutz, M., Zountsas, B., Garganis, K.,
  Rosenow, F., Hermsen, A., von Oertzen, T.J., Diepgen, T.L., Avanzini, G.,
  {EEBB Consortium}: Histopathological findings in brain tissue obtained during
  epilepsy surgery. N. Engl. J. Med.  \textbf{377}(17),  1648--1656 (Oct 2017)

\bibitem{David2021-zb}
David, B., Kr{\"o}ll-Seger, J., Schuch, F., Wagner, J., Wellmer, J., Woermann,
  F., Oehl, B., Van~Paesschen, W., Breyer, T., Becker, A., Vatter, H.,
  Hattingen, E., Urbach, H., Weber, B., Surges, R., Elger, C.E., Huppertz,
  H.J., R{\"u}ber, T.: External validation of automated focal cortical
  dysplasia detection using morphometric analysis. Epilepsia  (Feb 2021)

\bibitem{Fawaz2021-ag}
Fawaz, A., Williams, L.Z.J., Alansary, A., Bass, C., Gopinath, K., da~Silva,
  M., Dahan, S., Adamson, C., Alexander, B., Thompson, D., Ball, G.,
  Desrosiers, C., Lombaert, H., Rueckert, D., David~Edwards, A., Robinson,
  E.C.: Benchmarking geometric deep learning for cortical segmentation and
  neurodevelopmental phenotype prediction (Dec 2021)

\bibitem{Fischl2012-cx}
Fischl, B.: {FreeSurfer}. Neuroimage  \textbf{62}(2),  774--781 (Aug 2012)

\bibitem{Fortin2018-of}
Fortin, J.P., Cullen, N., Sheline, Y.I., Taylor, W.D., Aselcioglu, I., Cook,
  P.A., Adams, P., Cooper, C., Fava, M., McGrath, P.J., McInnis, M., Phillips,
  M.L., Trivedi, M.H., Weissman, M.M., Shinohara, R.T.: Harmonization of
  cortical thickness measurements across scanners and sites. Neuroimage
  \textbf{167},  104--120 (Feb 2018)

\bibitem{Gill2021-jf}
Gill, R.S., Lee, H.M., Caldairou, B., Hong, S.J., Barba, C., Deleo, F.,
  D'Incerti, L., Mendes~Coelho, V.C., Lenge, M., Semmelroch, M., Schrader,
  D.V., Bartolomei, F., Guye, M., Schulze-Bonhage, A., Urbach, H., Cho, K.H.,
  Cendes, F., Guerrini, R., Jackson, G., Hogan, R.E., Bernasconi, N.,
  Bernasconi, A.: Multicenter validation of a deep learning detection algorithm
  for focal cortical dysplasia. Neurology  (Sep 2021)

\bibitem{Gong2019-wg}
Gong, S., Chen, L., Bronstein, M., Zafeiriou, S.: {SpiralNet++}: A fast and
  highly efficient mesh convolution operator. In: 2019 {IEEE/CVF} International
  Conference on Computer Vision Workshop ({ICCVW}). pp.~0--0. IEEE (Oct 2019)

\bibitem{Isensee2021-yx}
Isensee, F., Jaeger, P.F., Kohl, S.A.A., Petersen, J., Maier-Hein, K.H.:
  {nnU-Net}: a self-configuring method for deep learning-based biomedical image
  segmentation. Nat. Methods  \textbf{18}(2),  203--211 (Feb 2021)

\bibitem{Ronneberger2015-hd}
Ronneberger, O., Fischer, P., Brox, T.: {U-Net}: Convolutional networks for
  biomedical image segmentation. In: Medical Image Computing and
  {Computer-Assisted} Intervention -- {MICCAI} 2015. pp. 234--241. Springer
  International Publishing (2015)

\bibitem{Spitzer2022}
Spitzer, H., Ripart, M., Whitaker, K., D'Arco, F., Mankad, K., Chen, A.A.,
  Napolitano, A., De~Palma, L., De~Benedictis, A., Foldes, S., Humphreys, Z.,
  Zhang, K., Hu, W., Mo, J., Likeman, M., Davies, S., G{\"u}ttler, C., Lenge,
  M., Cohen, N.T., Tang, Y., Wang, S., Chari, A., Tisdall, M., Bargallo, N.,
  Conde-Blanco, E., Pariente, J.C., Pascual-Diaz, S., Delgado-Mart{\'\i}nez,
  I., P{\'e}rez-Enr{\'\i}quez, C., Lagorio, I., Abela, E., Mullatti, N.,
  O'Muircheartaigh, J., Vecchiato, K., Liu, Y., Caligiuri, M.E., Sinclair, B.,
  Vivash, L., Willard, A., Kandasamy, J., McLellan, A., Sokol, D., Semmelroch,
  M., Kloster, A.G., Opheim, G., Ribeiro, L., Yasuda, C., Rossi-Espagnet, C.,
  Hamandi, K., Tietze, A., Barba, C., Guerrini, R., Gaillard, W.D., You, X.,
  Wang, I., Gonz{\'a}lez-Ortiz, S., Severino, M., Striano, P., Tortora, D.,
  K{\"a}lvi{\"a}inen, R., Gambardella, A., Labate, A., Desmond, P., Lui, E.,
  O'Brien, T., Shetty, J., Jackson, G., Duncan, J.S., Winston, G.P., Pinborg,
  L.H., Cendes, F., Theis, F.J., Shinohara, R.T., Cross, J.H., Baldeweg, T.,
  Adler, S., Wagstyl, K.: Interpretable surface-based detection of focal
  cortical dysplasias: a multi-centre epilepsy lesion detection study. Brain
  \textbf{145}(11),  3859--3871 (Nov 2022)

\bibitem{Wagner2011-ud}
Wagner, J., Urbach, H., Niehusmann, P., von Lehe, M., Elger, C.E., Wellmer, J.:
  Focal cortical dysplasia type {IIb}: completeness of cortical, not
  subcortical, resection is necessary for seizure freedom. Epilepsia
  \textbf{52}(8),  1418--1424 (Aug 2011)

\bibitem{Walger2023-ja}
Walger, L., Adler, S., Wagstyl, K., Henschel, L., David, B., Borger, V.,
  Hattingen, E., Vatter, H., Elger, C.E., Baldeweg, T., Rosenow, F., Urbach,
  H., Becker, A., Radbruch, A., Surges, R., Reuter, M., Cendes, F., Wang, Z.I.,
  Huppertz, H.J., R{\"u}ber, T.: Artificial intelligence for the detection of
  focal cortical dysplasia: Challenges in translating algorithms into clinical
  practice. Epilepsia  (Jan 2023)

\end{thebibliography}
%

\newpage
\section*{Supplementary Figures}
\renewcommand{\figurename}{Supplementary Fig.}
\setcounter{figure}{0}
\begin{figure}[ht]   
\includegraphics[width=\textwidth]{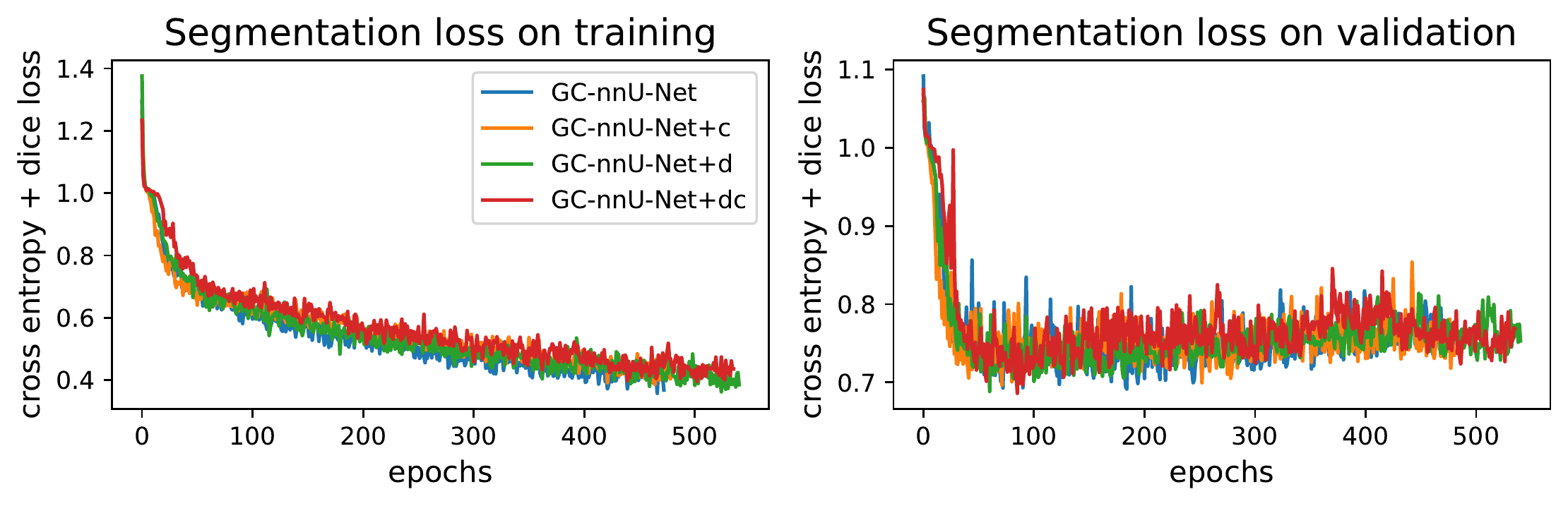}
\caption{
Training and validation segmentation losses for four experiments. All models were trained with early stopping to prevent overfitting. The model weights for the best performance on the validation loss are stored and training is stopped if no improvement in performance is seen after an additional 400 epochs. In the above examples, all models achieved peak performance on the validation loss at between 70 and 170 epochs.
}
\end{figure}

\begin{figure}[ht]
\includegraphics[width=\textwidth]{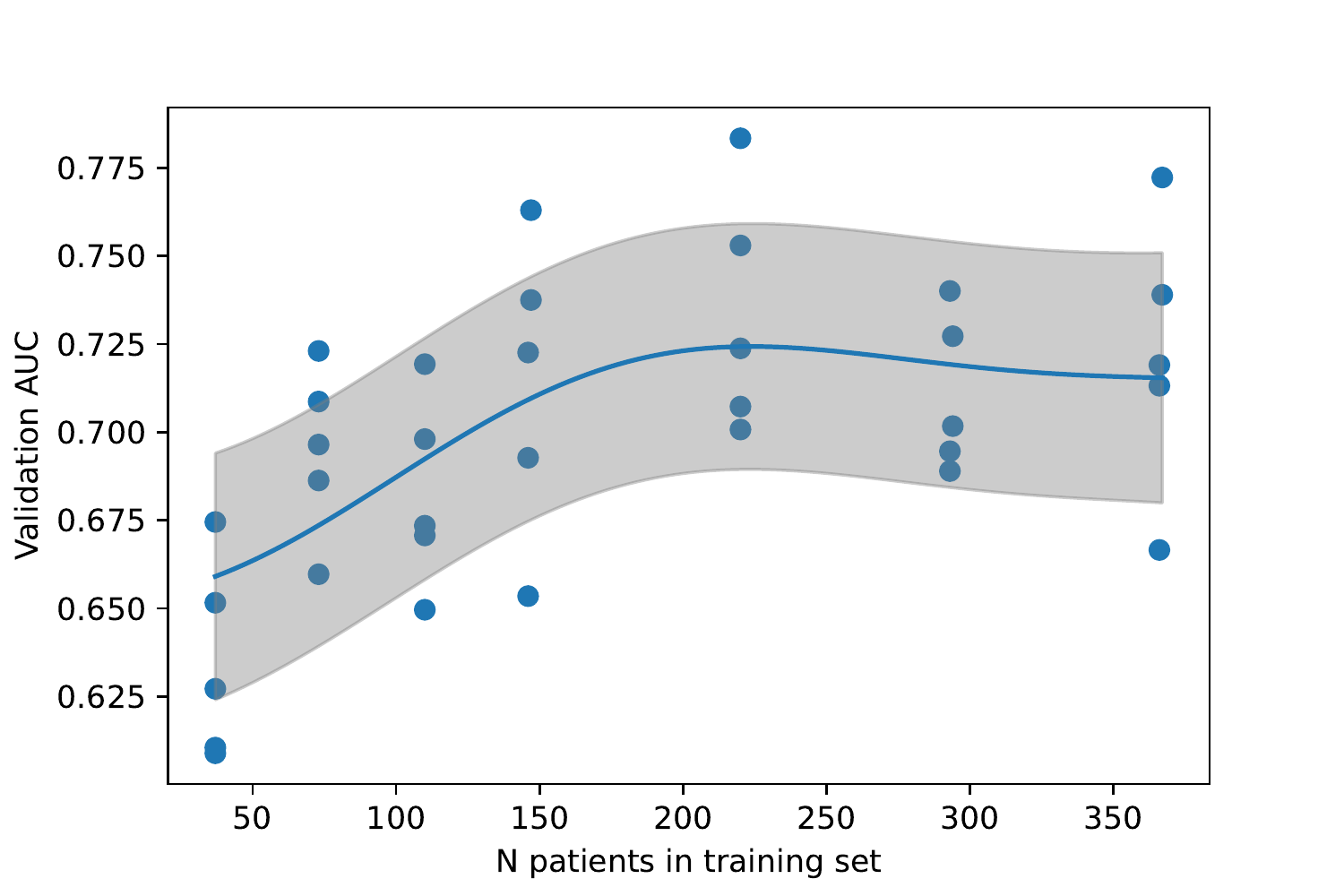}
\caption{
To assess the impact of sample size on model performance, models were retrained on subsamples of the training set at fixed fractions (0.1, 0.2, 0.3, 0.4, 0.6, 0.8, 1.0). Validation AUC was quantified for each model, across 5 folds. Blue line shows Gaussian process regression fit with standard deviation in gray. We see an improvement in performance which plateaued at around 220 subjects. 
}
\end{figure}

\end{document}